\documentclass[aps,prb,
  reprint, 
  tightenlines,
  citeautoscript,%
  showpacs,floatfix,
  superscriptaddress]{revtex4-1}

\usepackage{bm}
\usepackage{amssymb}
\usepackage{graphicx}
\usepackage{amsmath}
\usepackage{textcomp}
\usepackage{multirow}

\begin{document}
\title{ Nonperturbative model for optical response under intense
periodic fields with application to graphene in a strong
perpendicular magnetic field
}
\author{J. L. Cheng}
\affiliation{The Guo China-US Photonics Laboratory, Changchun
  Institute of Optics, fine Mechanics and Physics, Chinese Academy of Sciences, 3888 Eastern South Lake Road,
Changchun, Jilin 130033, China.}
\affiliation{University of Chinese Academy of Sciences, Beijing 100049, China}
\author{C. Guo}%
\affiliation{The Guo China-US Photonics Laboratory, Changchun Institute of Optics, fine Mechanics and
Physics, Chinese Academy of Sciences, 3888 Eastern South Lake Road,
Changchun, Jilin 130033, China.}
\affiliation{The Institute of Optics, University of Rochester, Rochester, NY 14627, USA.}
\date{\today}
\begin{abstract}
  Graphene exhibits extremely strong optical nonlinearity when
    a strong perpendicular magnetic field is applied, the response
    current shows strong field dependence even for moderate light
    intensity, and the perturbation theory fails. 
    We nonperturbatively calculate full optical conductivities induced
    by a periodic field in an equation-of-motion 
    framework based on the Floquet theorem,  with the scattering
    described phenomenologically. The nonlinear response at high fields
    is understood in terms of the dressed electronic 
    states, or Floquet states, which is  further
    characterized by the optical conductivity for a weak probe light
    field. This approach is illustrated for a
    magnetic field at $5$~T and a driving field with photon energy
    $0.05$~eV.  Our results show that the perturbation theory works
    only for weak 
    fields  $<3$~kV/cm, confirming the extremely  strong light matter
    interaction for Landau levels of graphene. This approach can be
    easily extended  
    to the calculation of optical conductivities in other systems.   
\end{abstract}
\maketitle

\section{Introduction}
Landau levels (LLs) of graphene show unique properties including a 
large cyclotron energy $\hbar\omega_c\approx 36
\sqrt{B(\text{Tesla})}$~meV and nonequidistant energies. This suggests
that graphene in a strong magnetic field should be a
good platform for demonstrating many fundamental dynamics concepts
\cite{Rev.Mod.Phys._83_1193_2011_Goerbig}, even at room
temperature. Recent studies of the nonlinear responses have been
extended to wavelengths in the infrared \cite{Phys.Rev.B_86_115427_2012_Rao,Phys.Rev.Lett._108_255503_2012_Yao,J.Phys.Condens.Matter_25_054203_2013_Yao,NanoLett._17_2184_2017_Koenig-Otto,Phys.Rev.B_96_045427_2017_Brem,arXiv:1711.04408}. A huge
optical susceptibility is predicted by Yao and Belyanin
\cite{Phys.Rev.Lett._108_255503_2012_Yao,J.Phys.Condens.Matter_25_054203_2013_Yao} 
and confirmed by the four wave mixing (FWM) experiments of 
K\"onig-Otto {\it et al.} in the far infrared
\cite{NanoLett._17_2184_2017_Koenig-Otto}. Proposed applications for
graphene-based photonics include the generation of 
entangled photons \cite{Phys.Rev.Lett._110_077404_2013_Tokman}, an
all-optical switch \cite{LaserPhys._27_16201_2017_Shiri}, tunable
lasers \cite{Phys.Rev.B_96_045427_2017_Brem}, the dynamic
control of coherent pulses \cite{Sci.Rep._7_2513_2017_Yang}, and the
demonstration of  optical bistability and optical multistability
\cite{Phys.B_497_67_2016_Solookinejad,
  J.Appl.Phys._117_183101_2015_Hamedi}.

Theoretically, optical nonlinearities are mostly studied in an
equation-of-motion framework, where solutions of the dynamical
equations can be obtained in the rotating wave approximation (RWA)
\cite{Phys.Rev.Lett._108_255503_2012_Yao,J.Phys.Condens.Matter_25_054203_2013_Yao}
or in perturbation method \cite{arXiv:1711.04408}. 
RWA is suitable for resonant transitions and modest incident laser intensities, which are usually discussed
between lowest several LLs. The perturbation theory can easily include the
contribution from all LLs, and it works well only for weak
light intensities. However, both theoretical prediction\cite{Phys.Rev.Lett._108_255503_2012_Yao} and experimental
  measurement\cite{NanoLett._17_2184_2017_Koenig-Otto} confirm that the LLs
  of graphene have a very weak saturation fields with values around a few kV/cm. For
  high fields, the optical  
response could be obtained by numerical simulation, but
often such calculations do not lead to physical insight into the
underlying physics. In this paper, we propose to investigate
  the nonlinear response in the basis of
  Floquet states.

When electrons are driven by a periodic field at frequency
  $\Omega$, the electronic states can be expressed by Floquet theorem
  \cite{Phys.Rev._138_B979_1965_Shirley} as Floquet states, which are
  nonperturbative solution of Schr\"odinger equation with light matter
  interaction.  This
    approach is used to study the gap opening by a laser field 
in graphene \cite{Phys.Rev.B_78_201406_2008_Lopez-Rodriguez,
  Appl.Phys.Lett._98_232103_2011_Calvo, Nat.Commun._6_7047_2015_Sentef}
and Floquet topological insulators
\cite{PhysStatusSolidiRRL_7_101_2013_Cayssol}. For graphene in the
absence of magnetic fields, it has also been applied to study the transport and linear optical properties 
\cite{Appl.Phys.Lett._97_11907_2010_Xu,
  Phys.Rev.B_83_245436_2011_Zhou,Phys.Rev.Lett._113_266801_2014_FoaTorres,
Phys.Rev.Lett._113_236803_2014_Kundu}, the dynamic Franz-Keldysh 
effect \cite{Phys.Rev.Lett._76_4576_1996_Jauho,Phys.Rev.B_83_245436_2011_Zhou}, and side band effects
\cite{Phys.Rev.B_83_245436_2011_Zhou}. Recently, Kibis {\it et
    al.}\cite{Phys.Rev.B_93_115420_2016_Kibis} used Floquet theorem to
  study the 
optical and transport effects of dressed LLs of graphene by a
monochromatic field. When there is adequate
damping, the system can reach a steady state that is also periodic in
time and can be probed\cite{Phys.Rev.B_71_115313_2005_Torres} by a weak light with a different frequency
  $\omega$. Generally, the response current includes components at
frequencies $l\Omega$  and $l\Omega+\omega$ with integer $l$. Most
studies focus on the response current components at frequencies
  $\Omega$ and  $\omega$; few discussion is performed for components
  at other 
  frequencies, which are essential quantities for many nonlinear
  optical phenomena including third harmonic generation (THG) and FWM.

In this paper we extend the Floquet theorem to study optical
nonlinearity in the equation-of-motion framework under relaxation time
approximation, and set up a connection between the obtained
expressions and the perturbation results. We apply this approach
to the optical response of LLs of graphene. Due to the strong light matter
interaction, this approach is illustrated for field below a
few tens kV/cm, which can be generated by continuous wave laser or long
duration laser pulse. For considered field strength, the relaxation time
approximation is still a widely used
description\cite{Phys.Rev.Lett._108_255503_2012_Yao} for
scattering. As such, we discuss the nonlinear response including THG
and FWM.

We organize the paper as follows. In Sec.~\ref{sec:model} we give
  all the expression for the response currents and conductivities from
  a general point of view; in Sec.~\ref{sec:gh} we
  apply the model to graphene under a strong perpendicular magnetic
  field, and show the optical nonlinearities for its steady state and
  the probe conductivities when a probe field is introduced; in Sec.~\ref{sec:con} we
  conclude and discuss the possible issue to be fixed in the future.

\section{Method\label{sec:model}}
We consider optical response of a $N$-level system (states labeled by 
Roman letters $n=1,2,\cdots,N$) to an electric
field $\bm E(t)$. The Hamiltonian can be written as
\begin{equation}
  \hat{H}(t)=\hat{H}_0 + e \theta(t)\bm E(t)\cdot\hat{\bm \xi}\,,
\end{equation}
where $-e$ is the electron charge,  $\hat H_0$ is the unperturbed Hamiltonian described by a
$N\times N$ matrix with elements $(\hat H_0)_{mn}=\varepsilon_m \delta_{mn}$,
and $\hat{\bm \xi}$ is a matrix describing the dipole interaction. A
quantity with a hat $\hat{O}$ stands for a matrix with row and column
indexed by the level index $n$. 
The electric field $\bm E(t) =\bm E_{\text{drv}}(t) + \bm
E_{\text{prb}}(t)$ includes a driving field $\bm E_{\text{drv}}(t)$,
which can be strong, and a probe field $\bm 
E_{\text{prb}}(t)$, which is usually very weak. The light-matter
interaction is turned on at $t=0$ suddenly.  The time evolution of
the system is 
described by the equation of motion
\begin{eqnarray}
  \hbar \frac{\partial \hat \rho(t)}{\partial t} = -i[\hat H(t), \hat
    \rho(t)] -\hbar \gamma [\hat \rho(t) - \hat \rho^0]\,, \label{eq:eom}
\end{eqnarray}
where $\hat \rho(t)$ is a single-particle density matrix. The last
term is a widely used phenomenological description of the scattering,  with $\hat \rho^0$ the density matrix at equilibrium
state and $\gamma$ a relaxation parameter. We organize the
formal solution as
\begin{eqnarray}
  \hat \rho(t)&=&\hat \rho^0+\hat \rho_{\text{drv}}(t) + \hat \rho_{\text{prb}}(t)\,,\\
  \hat \rho_{\text{drv}}(t) &=& \frac{e}{i\hbar}\int_{0}^t\!\!\!d\tau
  e^{\gamma(\tau-t)} \hat {\cal
    U}(t,\tau) \bm
  E_{\text{drv}}(\tau)\cdot[\hat {\bm\xi},\hat \rho^0]\hat {\cal U}(\tau,t)\,,\\
  \rho_{\text{prb}}(t)&=&\frac{e}{i\hbar} \int_{0}^t\!\!\!d\tau
  e^{\gamma(\tau-t)} \hat {\cal
    U}(t,\tau) \bm E_{\text{prb}}(\tau)\cdot[\hat{\bm\xi},\hat \rho(\tau)] \hat{\cal
    U}(\tau,t)\,,\quad\label{eq:rhop}
\end{eqnarray}
where $\hat{\cal U}(t,\tau)=\sum_\alpha
\psi_\alpha(t)\psi_\alpha^\dag(\tau)$ is an unitary matrix, and $\psi_{\alpha}(t)$ satisfies the Schr\"odinger equation
\begin{eqnarray}
  i\hbar\partial_t \psi_\alpha(t) = [\hat H_0 + e\bm E_d(t)\cdot\hat{\bm\xi}]
  \psi_\alpha(t)\,, \quad 
    \text{for }t>0\,. \label{eq:sch}
\end{eqnarray}
Because we are only interested in the solution at $t>0$, the
factor $\theta(t)$ appearing in the Hamiltonian $H(t)$ can be
ignored. Here the Greek subscript $\alpha$ stands for the index of the
eigenstate with the inclusion of the driving field. Obviously, the unitary matrix satisfies
$\hat{\cal U}(\tau,\tau) = I$.  We are interested in the response current density\cite{arXiv:1711.04408} $\bm
    J(t)=-e\text{Tr}[\hat{\bm v}\hat\rho(t)]$ with $\hat{\bm
      v} = [\hat{\bm\xi},\hat H(t)]/(i\hbar)=[\hat{\bm\xi},\hat H_0]/(i\hbar)$. Further we can write it
    as $\bm J(t)=\bm J_{\text{drv}}(t) + \bm
    J_{\text{prb}}(t)$, where the driving current is $\bm J_{\text{drv}}(t) = -e\text{Tr}[\hat{\bm v}\hat 
    \rho_{\text{drv}}(t)]$ and the probe current $\bm J_{\text{prb}}(t) =
    -e\text{Tr}[\hat{\bm v}\hat \rho_{\text{prb}}(t)]$.

Here we consider a special driving field, which is periodic
\begin{equation}
   E_{\text{drv}}^d(t)=\sum_l E_{\text{drv}}^{(l);d}e^{-il\Omega t}\,,
\end{equation}
where the Roman superscripts stand for the Cartersian directions
$\hat{\bm x}$ or $\hat{\bm y}$. 
Using the Floquet theorem \cite{Phys.Rev._138_B979_1965_Shirley}, the
eigen states are Floquet states, which are dressed electronic states
and can be expanded as
\begin{equation}
  \psi_\alpha(t)=e^{-i\epsilon_\alpha t/\hbar}\sum_l e^{-i l\Omega t}
u_{\alpha}^{(l)}\,,\label{eq:psi}
\end{equation}
where $\epsilon_\alpha$ is the quasi-energy,
  $u_\alpha^{(l)}$ is a $N$-row vector, and $\{u_{\alpha}^{(l)},l=\cdots,-1,0,1,\cdots\}$
forms the $\alpha$th eigen vectors. They
satisfy the eigen equation 
\begin{equation}
  (l\hbar \Omega + \epsilon_\alpha)u_{\alpha}^{(l)} =
  \hat{H}_0u_{\alpha}^{(l)} + \sum_n e E_{\text{drv}}^{(n);d}\hat{\xi}^d
  u_{\alpha}^{(l-n)} \,.\label{eq:quasi-eigenf}
\end{equation}
Although $\{u_{\alpha}^{(l+m)},l=\cdots,-1,0,1\cdots\}$ for any
integer $m$ is also an 
eigenstate of Eq.~(\ref{eq:quasi-eigenf}) with energy
$\epsilon_\alpha+m\hbar\Omega$, they correspond to
the same state $\psi_\alpha(t)$ of the Schr\"odinger equation
~(\ref{eq:sch}); only one of them needs to be
considered. The normalization of $\psi_\alpha(t)$ in
  Eq.~(\ref{eq:psi}) gives 
    $\text{Tr}[\hat{\cal
        A}_{\alpha_1\alpha_2}^{(l)}]=\delta_{\alpha_1\alpha_2}\delta_{l,0}$
    with
    $\hat{\cal A}_{\alpha_1\alpha_2}^{(l)}= \sum_{l_1}
    u_{\alpha_2}^{(l_1)}\left[u_{\alpha_1}^{(l_1-l)}\right]^\dag$. 
After some algebra, we get 
\begin{eqnarray}
  \hat \rho_{\text{drv}}(t) &=& \sum_{l}e^{-il\Omega t} \hat\rho_{\text{drv}}^{(l)}(t)\,,\\
 \hat\rho_{\text{drv}}^{(l)}(t) &=& \sum_{\alpha_1\alpha_2 l_1}{\cal A}_{\alpha_2\alpha_1}^{(l_1)} G_{\alpha_1\alpha_2}^{(l-l_1)} \notag\\
  && \times [1-e^{-\gamma t} e^{i(l-l_1)\Omega
      t}e^{-i(\epsilon_{\alpha_1}-\epsilon_{\alpha_2})t/\hbar}]\,,\\
      G_{\alpha_1\alpha_2}^{(l)} &=&
      \frac{e\sum_{l_2}E^{(l_2);d}_{\text{drv}}\text{Tr}\left\{
        [\hat\xi^d,\hat\rho^0]\hat{\cal
          A}_{\alpha_1\alpha_2}^{(l-l_2)}\right\}}{l\hbar\Omega
        - (\epsilon_{\alpha_1}-\epsilon_{\alpha_2}) + i\hbar\gamma}\,. \label{eq:gl}
\end{eqnarray}
Here $\hat\rho_{\text{drv}}^{(l)}(t)$ includes  oscillating terms related to
the correlations between Floquet states, but decaying with a factor 
$e^{-\gamma t}$. These
oscillations correspond to damped Rabi oscillations. As $t\to\infty$, 
they vanish, then $\hat\rho_{\text{drv}}^{(l)}(t)$ and
$\hat\rho_{\text{drv}}(t)$ reach their steady state, which are also periodic in time. In the clean limit $\gamma\to0$,
 an apparent divergence appears in the expression of
 $G^{(0)}_{\alpha\alpha}$, which can be shown to vanish from
 Eq.~(\ref{eq:quasi-eigenf}). \footnote{Equation~(\ref{eq:quasi-eigenf})
   gives 
    $\sum_{n_1n_2}eE^{(n_1);d}_{\text{drv}}\hat\xi^du_\alpha^{(n_2)}[u_\alpha^{(n_1+n_2)}]^\dag
    =\sum_l(l\hbar\Omega+\epsilon_\alpha-\hat
    H_0)u_\alpha^{(l)}[u_{\alpha}^{(l)}]^\dag$.  By substituting this
    expression into the numerator of 
    Eq.~(\ref{eq:gl}), it can be simplified to $[\hat H_0,\hat
      \rho_0]$, which is zero.}
 This is not
  surprising because our results are the full solutions of
  Schr\"odinger equation, which should not diverge. 
  The asymptotic current as 
$t\to\infty$ is
\begin{eqnarray}
  J_{\text{drv}}^d(t\to\infty) &=& \sum_{l} e^{-il\Omega t} J_{\text{drv}}^{(l);d}\,,\\
  J_{\text{drv}}^{(l);d} &=& -e\sum_{\alpha_1\alpha_2
    l_1}v^{(l_1);d}_{\alpha_2\alpha_1}
  G_{\alpha_1\alpha_2}^{(l-l_1)}\,. \label{eq:linearj}
\end{eqnarray}
Here we used notation $\text{Tr}[\hat X \hat{\cal
    A}_{\alpha_1\alpha_2}^{(l)}]=X_{\alpha_1\alpha_2}^{(l)}$ for $\hat
X=\hat v^d$.

The effects induced by the driving field can also be  detected by a
probe light $\bm E_{\text{prb}}(t)$. It leads to a change of the density
matrix by $\hat\rho_{\text{prb}}(t)$, and induces a probe current
  density $\bm J_{\text{prb}}(t)$. Up to the linear order
of $\bm E_{\text{prb}}$, we solve $\hat\rho_{\text{prb}}(t)$ in
Eq.~(\ref{eq:rhop}) by setting
$\hat\rho(\tau)=\hat\rho^0+\hat\rho_{\text{drv}}(\tau)$ and take 
the asymptotic results as $t\to\infty$ to give
\begin{eqnarray}
    \hat\rho_{\text{prb}}(t\to\infty)&=&\int\frac{d\omega}{2\pi} \sum_l
    e^{-il\Omega t -i\omega t} e E^b_p(\omega) \sum_{\alpha_1\alpha_2l_1}\hat{\cal
      A}^{(l_1)}_{\alpha_2\alpha_1} \notag\\
    && \times\left[{\cal G}_{\alpha_1\alpha_2}^{(l-l_1);b}(\omega)+
      {\cal P}^{(l-l_1);b}_{\alpha_1\alpha_2}(\omega) \right]\,,
\end{eqnarray}
with
\begin{eqnarray}
    {\cal G}_{\alpha_1\alpha_2}^{(l);b}(\omega)&=&\frac{\text{Tr}\left\{
        [\hat\xi^b,\hat\rho^0]\hat{\cal
          A}_{\alpha_1\alpha_2}^{(l)}\right\}}{l\hbar\Omega + \hbar\omega
      - (\epsilon_{\alpha_1}-\epsilon_{\alpha_2}) + i\hbar\gamma}\,,\label{eq:sigmaprb1}\\
    {\cal P}^{(l);b}_{\alpha_1\alpha_2}(\omega) &=&
    \frac{\sum_{\alpha l_2}
      \left[\xi^{(l-l_2);b}_{\alpha_1\alpha}{G}_{\alpha\alpha_2}^{(l_2)}-{G}_{\alpha_1\alpha}^{(l_2)}\xi^{(l-l_2);b}_{\alpha\alpha_2}\right]}{l\hbar\Omega+\hbar\omega-(\epsilon_{\alpha_1}-\epsilon_{\alpha_2})+i\hbar\gamma}\,. \label{eq:sigmaprb}
\end{eqnarray}
Here the term ${\cal G}$ is from $\hat\rho^0$, and the term ${\cal
  P}$ is from $\hat\rho_{\text{drv}}$.  The current density is then
\begin{equation*}
  J^d_{\text{prb}}(t\to\infty) = \int\frac{d\omega}{2\pi} \sum_le^{-i(l\Omega +
    \omega)t}\sigma_{\text{prb}}^{(l);db}(\omega)E_{\text{prb}}^b(\omega) \,,
\end{equation*}
with the probe conductivity
\begin{equation}
  \sigma_{\text{prb}}^{(l);db}(\omega)= -e^2\sum_{\alpha_1\alpha_2 l_1}
  v^{(l-l_1);d}_{\alpha_2\alpha_1}\left[{\cal G}^{(l_1);b}_{\alpha_1\alpha_2}(\omega)+{\cal P}^{(l_1);b}_{\alpha_1\alpha_2}(\omega)\right]\,.\label{eq:sigmaprobe}
\end{equation}
The quantities $J^{(l);d}_{\text{drv}}$ and
    $J^{(l);d}_{\text{prb}}(\omega)$ are experimental observable
    quantities, which can be extracted by measuring the light
    intensity of the electromagnetic radiation at frequencies
    $l\Omega$ and $l\Omega+\omega$, respectively. 
The driving field affects the probe conductivities in two
    aspects: One is the energy spectrum of quasi-states, which appear
    in the denominators of Eqs.~(\ref{eq:sigmaprb1}) and
    (\ref{eq:sigmaprb}). In the limit of weak 
    relaxation, this contribution is significant around resonant
    peaks. The other is that the steady states of density matrix,
    including both the occupations at each quasi-state and the
    polarization between them, are changed by the driving field. The
    latter dominates the cases away
    from the resonance. 

It is constructive to connect our results with
the usual perturbative conductivities. Here the order $l$ in both
$J^{(l);d}_{\text{drv}}$ and $\sigma_{\text{prb}}^{(l);db}(\omega)$
correspond to the response frequency $l\Omega$ and $l\Omega+\omega$,
respectively, instead of the orders of
the driving field. At a weak field,
Eq.~(\ref{eq:quasi-eigenf}) can be solved perturbatively by treating
the last term as a perturbation. In the lowest order with taking
$\epsilon_\alpha\to \epsilon_n = \varepsilon_n$ and
$X^{(l)}_{\alpha_1\alpha_2}\to X^{(l)}_{n_1n_2} = X_{n_1n_2} \delta_{l,0}$,
$J^{(1);d}_{\text{drv}}$  reduces to the perturbation results
\cite{arXiv:1711.04408}. In general, the remarkable differences come
from  the denominator of Eqs.~(\ref{eq:gl}) and 
(\ref{eq:sigmaprb}).  In the 
limit $\gamma\to0$, the absorption edge can be changed by the field, which leads to the so called dynamic Franz-Keldysh effects 
\cite{Phys.Rev.Lett._76_4576_1996_Jauho}. For a finite $\gamma$, in
the regime where perturbation theory works, this phenomenon may be
smeared out. Usually, the conductivities with the contribution from
$l\hbar\Omega$ may be discussed in the content of side-band effects. As
we will show later, they can be associated with the nonlinear responses for
weak driving field. 

\section{Results\label{sec:gh}}
We apply this approach to graphene in a strong magnetic field, 
$\bm B=B\hat{\bm z}$. The electronic states are LLs noted as 
$|\nu s n k\rangle$, where $\nu=+$ ($-$) is a valley index for the
$\bm K$ ($\bm K^\prime$) valley, $s=\pm$ is a
band index, $n\ge (1+\nu s)/2$ is a Landau index, and $k$ is a
continuum index. The eigen
energies are $E_{\nu s n}=s\varepsilon_n$, where with 
$\varepsilon_n=\sqrt{n}\hbar\omega_c$ with
$\hbar\omega_c=\sqrt{2}\hbar v_F/l_c$ and the magnetic length
$l_c=\sqrt{\hbar/(eB)}$, which depends on the index ``$sn$'' only. The
continuum index $k$ gives a degeneracy  ${\cal D}=g_s/(2\pi l_c^2)$
with $g_s=2$ for spin degeneracy, and this index will be suppressed
hereafter. There is no coupling between the
two valleys, and the matrix elements of position and
velocity operators in the $\nu$th valley can be written as $\bm\xi_{\nu;s_1n_1,s_2n_2} =
\sum_{\tau=\pm}\xi_{\nu;s_1n_1,s_2n_2}^\tau (\hat{\bm x}-i\tau
\hat{\bm y})/\sqrt{2}$ and $\bm
v_{\nu;s_1n_1,s_2n_2}=i\hbar^{-1}(s_1\varepsilon_{n_1}-s_2\varepsilon_{n_2})\bm\xi_{\nu;s_1n_1,s_2n_2}$.
Inversion symmetry connects the quantities in the two 
valleys according to $\bm\xi_{+;s_1n_1,s_2n_2} =
s_1s_2\bm\xi_{-;(-s_1)n_1,(-s_2)n_2}$. Considering the Hermiticity of
these quantities, all relevant matrix elements can be generated from
$v^+_{+;s_1n_1,s_2n_2}=s_1 v_F\delta_{n_1,n_2+1}(\delta_{n_2\neq
  0}/\sqrt{2} + \delta_{n_2,0}\delta_{s_2,-1})$.

In our calculations, the parameters are taken as $B=5$~T,
$\hbar\Omega=0.05$~eV, the chemical potential $\mu=0$~eV, the
temperature $T=10$~K, and $\hbar\gamma=10$~meV. The driving field is
$\bm E_{\text{drv}}^{(l)} = E_0\hat{\bm x}
(\delta_{l,1}+\delta_{l,-1})$, and the density matrix at equilibrium
is $\rho^0_{\nu;s_1n_1,s_2n_2}=[1+e^{(s_1\epsilon_{n_1}-\mu)/(k_BT)}]^{-1}\delta_{s_1s_2}\delta_{n_1n_2}$ 
with $k_B$ the Boltzmann constant. The calculated Floquet states
$\psi_\alpha(t)$ in the $\nu$ valley is denoted as $|\nu\alpha\rangle_f$ with
$\alpha=-N_c,-N_c+1,\cdots,N_c-1,N_c$ and $N_c= 20$ being the cutoff of
Landau index, and their quasi-energies are noted as $\epsilon_{\nu \alpha}$. The driving field is taken as $E_0<60$~kV/cm, in which our
results are converged for the specified $N_c$.  The energies of the lowest
several LLs are $\varepsilon_n=0$, $81$, and $115$~meV for $n=0$, $1$,
$2$, respectively. The driving photon energy does not match any
of the resonant conditions. 

For a linearly polarized field, the system retains the electron-hole 
symmetry, and thus we can choose the quasi-energy of the Floquet
states to satisfy $\epsilon_{\nu\alpha}=-\epsilon_{\nu (-\alpha)}$, and
$\epsilon_{\nu0}=0$; the occupation
$[\hat\rho^{(l)}_{\text{drv}}]_{\nu;sn,sn}$  at the LL $|\nu s
n\rangle$ also satisfies
$[\hat\rho^{(l)}_{\text{drv}}]_{\nu;+n,+n}=-[\hat\rho^{(l)}_{\text{drv}}]_{\nu;-n,-n}$
and $[\hat\rho^{(l)}_{\text{drv}}]_{\nu;-0,-0}=0$.  Furthermore, due
to the crystal symmetry, the current responses $J^{(l);d}_{\text{drv}}$ and
$\sigma_{\text{prb}}^{(l);db}(\omega)$ are nonzero only for odd order
of $l$, and the density matrix $\rho_{\text{drv}}^{(l)}(t)$ is nonzero
only for even order of $l$. For a comparison with our previous
work\cite{arXiv:1711.04408}, we denote the perturbative conductivities
as  $\sigma_{\text{pert}}^{(n)}$.  In this paper, the relevant
conductivities are $\sigma_{\text{pert}}^{(1);xx}(\Omega)$ and
$\sigma_{\text{pert}}^{(3);xxxx}(\Omega,\Omega,\pm \Omega)$ for the
driving field, and
$\sigma_{\text{pert}}^{(3);xxxx}(\Omega,\pm\Omega,\omega)$ for the probe field.

\subsection{Current density response to the driving field}
\begin{figure}[!htpb]
  \centering
  \includegraphics[width=7cm]{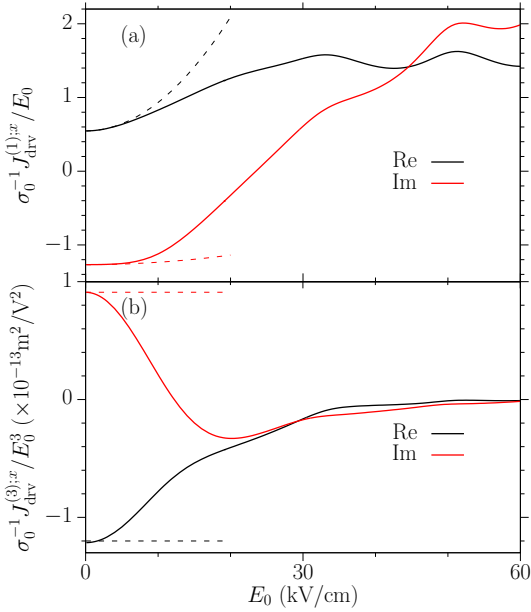}
  \caption{(color online) Current density induced by the periodic driving
    field for $E_0\le60$~kV/cm.
    (a) The effective linear  conductivity. (b) The effective
    conductivity for THG. The dashed curves are 
     perturbation results  given in the right hand side of
     Eqs.~(\ref{eq:pert1}) and (\ref{eq:pert2}).  
  } 
  \label{fig:jdrv}
\end{figure}
In Fig.~\ref{fig:jdrv} we plot the effective conductivity
$\sigma^{(1)}_{\text{eff}} = J^{(1);x}_{\text{drv}}/E_0$ at
fundamental frequency $\Omega$ and $\sigma^{(3)}_{\text{eff}} =
J^{(3);x}_{\text{drv}}/E_0^3$ at the third harmonic frequency $3\Omega$ as
 a function of the field amplitude $E_0$. 
We first compare these results with perturbation theory to determine
the field threshold. At weak field, up to the third order
conductivities the effective conductivities are expanded as 
\begin{eqnarray}
  \sigma^{(1)}_{\text{eff}} &\approx&
  \sigma_{\text{pert}}^{(1);xx}(\Omega) + 3
  \sigma_{\text{pert}}^{(3);xxxx}(\Omega,\Omega,-\Omega) E_0^2\,,\label{eq:pert1}\\
  \sigma^{(3)}_{\text{eff}}&\approx& \sigma_{\text{pert}}^{(3);xxxx}(\Omega,\Omega,\Omega)\,.\label{eq:pert2}
\end{eqnarray}
The perturbation results give $\sigma^{-1}_0
\sigma_{\text{pert}}^{(1);xx}=0.55- 1.27i$ with
$\sigma_0=e^2/(4\hbar)$, 
$\sigma_0^{-1}\sigma_{\text{pert}}^{(3);xxxx}(\Omega,\Omega,-\Omega)=(0.128 + 0.0108i) \times
10^{-12}$~m$^2$/V$^2$, and $\sigma_0^{-1}\sigma_{\text{pert}}^{(3);xxxx}(\Omega,\Omega,\Omega)=(-1.2 +0.91 i)\times
10^{-13}$~m$^2$/V$^2$; they are plotted in
Fig.~\ref{fig:jdrv} (a) as dashed curves. The perturbation results
agree with the full calculation very well when the field is
$E_0<5$~kV/cm (for $\sigma^{(1)}_{\text{eff}}$) or $E_0<3$~kV/cm (for
$\sigma^{(3)}_{\text{eff}}$) with an error less than
$5\%$. These small thresholds indicate extremely
strong interaction between the periodic field and the LLs of graphene. 
For large $E_0$, the real part of
$\sigma^{(1)}_{\text{eff}}$ increases with the field with a slope
slower than the perturbation results, and reaches a peak with value
$1.58\sigma_0$ at $E_0\sim 33$~kV/cm; then it shows one oscillation
and arrives at another peak at
$E_0\sim 51$~kV/cm. The imaginary part of $\sigma^{(1)}_{\text{eff}}$
firstly increases with the field by a larger slope, reaches a peak
around $E_0\sim51$~kV/cm. At $E_0\sim 33$~kV/cm, the imaginary part
shows a shoulder-like fine structure.  From our discussion in
Sec.~\ref{sec:model}, we can understand these features from the
properties of Floquet states. 

\begin{figure}[!htpb]
  \centering
  \includegraphics[width=7.5cm]{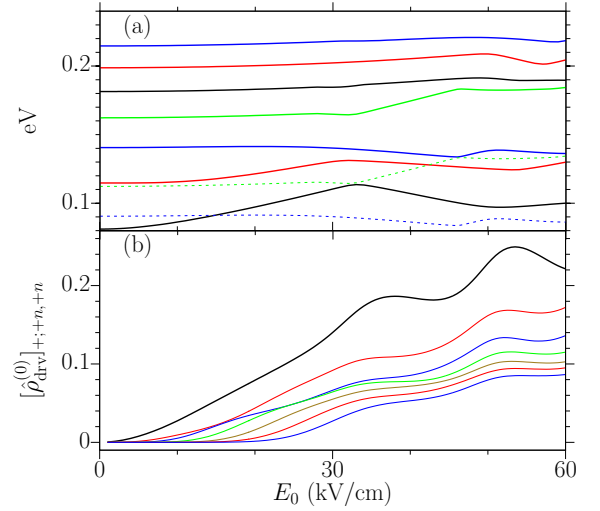}
  \caption{(color online) (a) Field dependence of quasi-energies
    $\epsilon_{+\alpha}$  in the $\bm K$ valley for
    $1\le\alpha\le7$. The two dashed curves corresponds to the energy
    $\epsilon_{+\alpha}-\hbar\Omega$  for $\alpha=3$ and $4$. (b)Field
    dependence of the zeroth order occupations at different LLs $|++n\rangle$
    for $1\le n\le 7$.}
  \label{fig:enrho}
\end{figure}
In Fig.~\ref{fig:enrho} (a) we plot the field dependence of the
quasi-energies $\epsilon_{+\alpha}$ for states $|+\alpha\rangle_f$ in
the $\bm K$ valley for $1\le\alpha\le7$.
 At zero field, these states correspond to the LLs $|++n\rangle$ for
$1\le n\le 7$. From electron-hole symmetry we can obtain their opposite
 energy counterparts $\epsilon_{+\alpha}=-\epsilon_{+(-\alpha)}$ and
 $\epsilon_{+0}=0$. We 
 focus on the quasi-state $|+1\rangle_f$, which corresponds 
to the LL $|++1\rangle$ at zero field. Its quasi-energy shows a complicated
dependence of $E_0$. It starts with 81~meV at zero field, and reaches
a local maximum around $113.5$~meV at about $E_0=33$~kV/cm, then
decreases to a local minimum with values $97$~meV at $E_0=51$~kV/cm, and
increases again. For small field, the energy corrections come  mostly
from the LLs $|+-0\rangle$ and $|+ s2\rangle$, due to the selection
rules.   

For stronger field, the Floquet states mix more LLs; the selection
rules between Floquet states can be greatly modified from those
between LLs. As an example, we analyze the behavior of the state
  $|+1\rangle_f$ around $E_0\sim 33$~kV/cm. The energy of this
Floquet state is 
close to that of $|+4\rangle_f$, which is shown in the same diagram by
plotting an equivalent quasi-energy $\epsilon_{+4}-\hbar\Omega$ as a
dashed curve. Their interaction is allowed and leads to an
anti-crossing (about 1~meV splitting).  Similar behavior occurs around 
the local minimum at $E_0\sim 51$~kV/cm, which is induced by the
interaction between the quasi-states $|+1\rangle_f$ and $|+3\rangle_f$.
Besides the modification of the selection rules, the strong  field can 
also greatly change the occupations on each LL, as shown in
Figure~\ref{fig:enrho} (b) for the occupation 
$[\hat\rho_{\text{drv}}^{(0)}(t\to\infty)]_{+;+n,+n}$ of the LL  
$|+sn\rangle$ for $1\le n\le 7$. When $E_0>30$~kV/cm, the occupation
at the LL $|++1\rangle$ is about 0.2, significantly deviating from its
thermal equilibrium ($\sim0$). Therefore,  both the
  quasi-energies and the populations show similar tendencies as the
  optical conductivity $\sigma^{(1)}_{\text{eff}}$, and they dominate
  the optical response induced by the driven field, as we discussed in
  Sec.~\ref{sec:model}. This partly explains why the
perturbation theory based on the thermal equilibrium fails.  

In Fig.~\ref{fig:jdrv} (b) we give the field dependence of the
  optical conductivity $\sigma^{(3)}_{\text{eff}}$ for THG. Both the
real and imaginary parts of $\sigma^{(3)}_{\text{eff}}$ decrease
quickly to very small values, and the imaginary part shows a
valley around $E_0\sim 20$~kV/cm. Similar to the
$\sigma^{(1)}_{\text{eff}}$, $\sigma^{(3)}_{\text{eff}}$ are mainly
affected by the changes of optically excited populations. Because
there is no specific physical process to distinguish its real and
imaginary parts, they behave in a similar way.

\subsection{Probe conductivities}
\begin{figure}[!htpb]
  \centering
  \includegraphics[width=7.5cm]{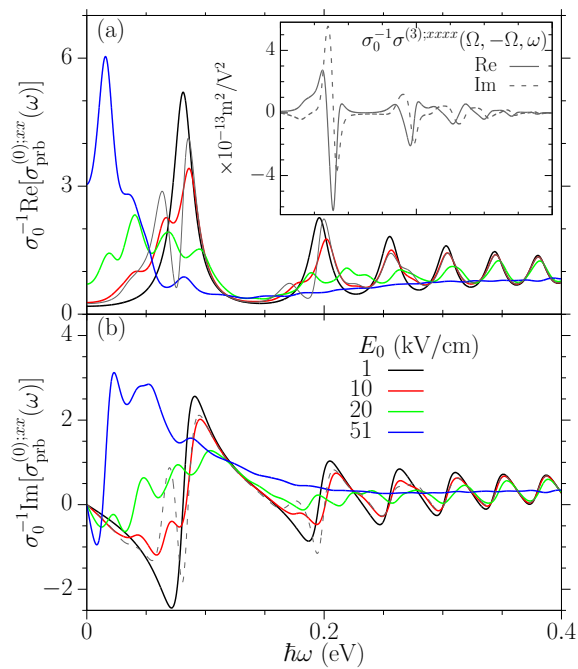}
  \caption{(color online) The spectrum of the probe conductivity $\sigma^{(0);xx}_{\text{prb}}(\omega)$ for
    different driving fields with (a) the real part
    (b) the imaginary part. The
    inset in (a) shows the perturbative third order conductivity\cite{arXiv:1711.04408}
    $\sigma^{(3);xxxx}(\Omega,-\Omega,\omega)$ with $x$-axis also in
    $\hbar\omega\in[0,0.4]$~eV.  The gray curves in
    (a) and (b) are the perturbative probe conductivity up to the
    third order for $E_0=10$~kV/cm. }  
  \label{fig:probexx0}
\end{figure}
The modification of the intense field on the LLs can also be probed by
a weak optical field. Similarly, for a very weak driving field, the probe
conductivities in Eq.~(\ref{eq:sigmaprobe}) can be approximated from perturbation theory, and up to the third
order we have
\begin{eqnarray}
  \sigma^{(0);xx}_{\text{prb}}(\omega) &\approx&
  \sigma_{\text{pert}}^{(1);xx}(\omega) + 6
  \sigma_{\text{pert}}^{(3);xxxx}(\Omega,-\Omega,\omega) E_0^2\,, \quad
\end{eqnarray}
Here $\sigma_{\text{pert}}^{(1);xx}(\omega)$ and
$\sigma_{\text{pert}}^{(3);xxxx}(\Omega,-\Omega,\omega)$ are the
perturbative conductivities calculated from our previous work \cite{arXiv:1711.04408}. In Fig.~\ref{fig:probexx0} we give the probe
conductivity $\sigma^{(0);xx}_{\text{prb}}(\omega)$ for a probe frequency
$\omega$. The calculation at weak field
$E_0=1$~kV/cm agrees with the perturbation results very well, and the real
part show many absorption peaks due to the transition between
different LLs. At $E_0=10$~kV/cm, the full calculation (red curves) and the
perturbation theory (gray curves) obviously differ around the first two
peaks. This is not surprising because the
    perturbation theory fails for field stronger than $3$~kV/cm. 
When the driving field is increased to
$E_0=20$~kV/cm, the full calculation of the probe conductivity does not
have any similarity to the perturbation results, and all peaks are
greatly smeared out. This is consistent 
with the change of the selection rules and the occupations of LLs,
which leads to a complicated behavior.  For $E_0=51$~kV/cm, most of
these peaks disappear, and a peak at very low frequency
emerges. The spectra look
    very similar to the optical conductivity for a doped graphene at
    high temperature, indicating a fully thermalization of the
    electrons by the driving field. 
Because this system
includes many energy scales, the probe
conductivities do not depend on the driving field in a simple way, and
their peaks can not be simply understood by side band effects.

\begin{figure}[!htpb]
  \centering
  \includegraphics[width=7.cm]{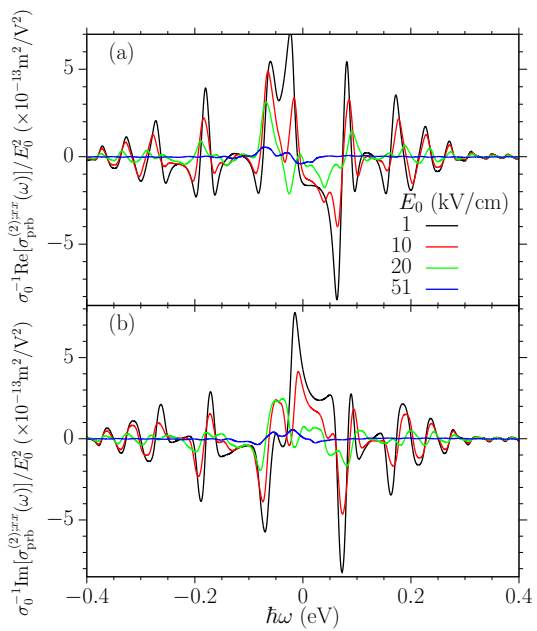}
  \caption{(color online) The spectrum of the probe conductivity
    $\sigma^{(0);xx}_{\text{prb}}(\omega)$ for
    different driving fields with (a) the real part
    (b) the imaginary part.}  
  \label{fig:probexx2}
\end{figure}
The detection can also be made for frequencies $\omega\pm 2\Omega$, which correspond to the
conductivity $\sigma^{(\pm2);xx}_{\text{prb}}(\omega)$ for FWM. The
results are shown in Fig.~\ref{fig:probexx2}. At weak driving fields,
they recover the perturbative conductivities 
$\sigma^{(2);xx}_{\text{prb}}(\omega)=\sigma^{(3);xxxx}(\Omega,\Omega,\omega)$
and
$\sigma^{(-2);xx}_{\text{prb}}(\omega)=[\sigma^{(3);xxxx}(\Omega,\Omega,-\omega)]^\ast$. The
perturbative results agree very well with the conductivities at
$E_0=1$~kV/cm. With increasing the field strength, the probe
conductivities differ from the perturbative ones obviously, which are
induced by the influences of the driving field on the system. 

\section{Conclusion\label{sec:con}}
In this study of the optical response induced by an intense periodic
field, we constructed a theoretical framework based on the Floquet
theorem, and derived the expressions for the full induced optical
current.  These expressions were used to study graphene subject to a
strong perpendicular magnetic field.
 By comparing with a perturbation theory
up to the third order, we determined the threshold field where
the perturbation theory broke down. We understood these
nonperturbative behavior from the Floquet states, which could be
detected by a weak light field in an optical method. Our results can
be extended to other systems.

There exist two unsolved issues in this approach: one is related to the
driving field. Because most strong incident fields are
laser pulses, they cannot be treated by a formalism based on fully
periodic fields in a straightforward way. It would be necessary to
extend the Floquet theorem to pulsed fields, even if appropriate
approximations were required. The other is related to the phenomenological
relaxation time approximation used in Eq.~(\ref{eq:eom}). As a widely
adopted approximation in perturbation theory for preliminary studies,
it is not clear whether or not it can be used, or how it would be
implemented for very strong fields. Although a  microscopic
treatment of the scattering is possible
\cite{Phys.Rev.B_96_045427_2017_Brem,Phys.Rev.B_75_035307_2007_Jiang}, 
it would still be desirable to develop simpler descriptions that might
lead to more physical insight. 

\acknowledgements
This work has been supported by CIOMP Y63032G160, CAS
QYZDB-SSW-SYS038, and NSFC 11774340. J.L.C acknowledges valuable
discussions with Prof. K. Shen and Prof. J.E. Sipe. 

%


\begin{thebibliography}{26}%
\makeatletter
\providecommand \@ifxundefined [1]{%
 \@ifx{#1\undefined}
}%
\providecommand \@ifnum [1]{%
 \ifnum #1\expandafter \@firstoftwo
 \else \expandafter \@secondoftwo
 \fi
}%
\providecommand \@ifx [1]{%
 \ifx #1\expandafter \@firstoftwo
 \else \expandafter \@secondoftwo
 \fi
}%
\providecommand \natexlab [1]{#1}%
\providecommand \enquote  [1]{``#1''}%
\providecommand \bibnamefont  [1]{#1}%
\providecommand \bibfnamefont [1]{#1}%
\providecommand \citenamefont [1]{#1}%
\providecommand \href@noop [0]{\@secondoftwo}%
\providecommand \href [0]{\begingroup \@sanitize@url \@href}%
\providecommand \@href[1]{\@@startlink{#1}\@@href}%
\providecommand \@@href[1]{\endgroup#1\@@endlink}%
\providecommand \@sanitize@url [0]{\catcode `\\12\catcode `\$12\catcode
  `\&12\catcode `\#12\catcode `\^12\catcode `\_12\catcode `\%12\relax}%
\providecommand \@@startlink[1]{}%
\providecommand \@@endlink[0]{}%
\providecommand \url  [0]{\begingroup\@sanitize@url \@url }%
\providecommand \@url [1]{\endgroup\@href {#1}{\urlprefix }}%
\providecommand \urlprefix  [0]{URL }%
\providecommand \Eprint [0]{\href }%
\providecommand \doibase [0]{http://dx.doi.org/}%
\providecommand \selectlanguage [0]{\@gobble}%
\providecommand \bibinfo  [0]{\@secondoftwo}%
\providecommand \bibfield  [0]{\@secondoftwo}%
\providecommand \translation [1]{[#1]}%
\providecommand \BibitemOpen [0]{}%
\providecommand \bibitemStop [0]{}%
\providecommand \bibitemNoStop [0]{.\EOS\space}%
\providecommand \EOS [0]{\spacefactor3000\relax}%
\providecommand \BibitemShut  [1]{\csname bibitem#1\endcsname}%
\let\auto@bib@innerbib\@empty
\bibitem [{\citenamefont {Goerbig}(2011)}]{Rev.Mod.Phys._83_1193_2011_Goerbig}%
  \BibitemOpen
  \bibfield  {author} {\bibinfo {author} {\bibfnamefont {M.~O.}\ \bibnamefont
  {Goerbig}},\ }\href {\doibase 10.1103/RevModPhys.83.1193} {\bibfield
  {journal} {\bibinfo  {journal} {Rev. Mod. Phys.}\ }\textbf {\bibinfo {volume}
  {83}},\ \bibinfo {pages} {1193} (\bibinfo {year} {2011})}\BibitemShut
  {NoStop}%
\bibitem [{\citenamefont {Rao}\ and\ \citenamefont
  {Sipe}(2012)}]{Phys.Rev.B_86_115427_2012_Rao}%
  \BibitemOpen
  \bibfield  {author} {\bibinfo {author} {\bibfnamefont {K.~M.}\ \bibnamefont
  {Rao}}\ and\ \bibinfo {author} {\bibfnamefont {J.~E.}\ \bibnamefont {Sipe}},\
  }\href {\doibase 10.1103/physrevb.86.115427} {\bibfield  {journal} {\bibinfo
  {journal} {Phys. Rev. B}\ }\textbf {\bibinfo {volume} {86}},\ \bibinfo
  {pages} {115427} (\bibinfo {year} {2012})}\BibitemShut {NoStop}%
\bibitem [{\citenamefont {Yao}\ and\ \citenamefont
  {Belyanin}(2012)}]{Phys.Rev.Lett._108_255503_2012_Yao}%
  \BibitemOpen
  \bibfield  {author} {\bibinfo {author} {\bibfnamefont {X.}~\bibnamefont
  {Yao}}\ and\ \bibinfo {author} {\bibfnamefont {A.}~\bibnamefont {Belyanin}},\
  }\href {\doibase 10.1103/PhysRevLett.108.255503} {\bibfield  {journal}
  {\bibinfo  {journal} {Phys. Rev. Lett.}\ }\textbf {\bibinfo {volume} {108}},\
  \bibinfo {pages} {255503} (\bibinfo {year} {2012})}\BibitemShut {NoStop}%
\bibitem [{\citenamefont {Yao}\ and\ \citenamefont
  {Belyanin}(2013)}]{J.Phys.Condens.Matter_25_054203_2013_Yao}%
  \BibitemOpen
  \bibfield  {author} {\bibinfo {author} {\bibfnamefont {X.}~\bibnamefont
  {Yao}}\ and\ \bibinfo {author} {\bibfnamefont {A.}~\bibnamefont {Belyanin}},\
  }\href {http://stacks.iop.org/0953-8984/25/i=5/a=054203} {\bibfield
  {journal} {\bibinfo  {journal} {J. Phys. Condens. Matter}\ }\textbf {\bibinfo
  {volume} {25}},\ \bibinfo {pages} {054203} (\bibinfo {year}
  {2013})}\BibitemShut {NoStop}%
\bibitem [{\citenamefont {K\"onig-Otto}\ \emph {et~al.}(2017)\citenamefont
  {K\"onig-Otto}, \citenamefont {Wang}, \citenamefont {Belyanin}, \citenamefont
  {Berger}, \citenamefont {de~Heer}, \citenamefont {Orlita}, \citenamefont
  {Pashkin}, \citenamefont {Schneider}, \citenamefont {Helm},\ and\
  \citenamefont {Winnerl}}]{NanoLett._17_2184_2017_Koenig-Otto}%
  \BibitemOpen
  \bibfield  {author} {\bibinfo {author} {\bibfnamefont {J.~C.}\ \bibnamefont
  {K\"onig-Otto}}, \bibinfo {author} {\bibfnamefont {Y.}~\bibnamefont {Wang}},
  \bibinfo {author} {\bibfnamefont {A.}~\bibnamefont {Belyanin}}, \bibinfo
  {author} {\bibfnamefont {C.}~\bibnamefont {Berger}}, \bibinfo {author}
  {\bibfnamefont {W.~A.}\ \bibnamefont {de~Heer}}, \bibinfo {author}
  {\bibfnamefont {M.}~\bibnamefont {Orlita}}, \bibinfo {author} {\bibfnamefont
  {A.}~\bibnamefont {Pashkin}}, \bibinfo {author} {\bibfnamefont
  {H.}~\bibnamefont {Schneider}}, \bibinfo {author} {\bibfnamefont
  {M.}~\bibnamefont {Helm}}, \ and\ \bibinfo {author} {\bibfnamefont
  {S.}~\bibnamefont {Winnerl}},\ }\href {\doibase 10.1021/acs.nanolett.6b04665}
  {\bibfield  {journal} {\bibinfo  {journal} {Nano Lett.}\ }\textbf {\bibinfo
  {volume} {17}},\ \bibinfo {pages} {2184} (\bibinfo {year}
  {2017})}\BibitemShut {NoStop}%
\bibitem [{\citenamefont {Brem}\ \emph {et~al.}(2017)\citenamefont {Brem},
  \citenamefont {Wendler},\ and\ \citenamefont
  {Malic}}]{Phys.Rev.B_96_045427_2017_Brem}%
  \BibitemOpen
  \bibfield  {author} {\bibinfo {author} {\bibfnamefont {S.}~\bibnamefont
  {Brem}}, \bibinfo {author} {\bibfnamefont {F.}~\bibnamefont {Wendler}}, \
  and\ \bibinfo {author} {\bibfnamefont {E.}~\bibnamefont {Malic}},\ }\href
  {\doibase 10.1103/PhysRevB.96.045427} {\bibfield  {journal} {\bibinfo
  {journal} {Phys. Rev. B}\ }\textbf {\bibinfo {volume} {96}},\ \bibinfo
  {pages} {045427} (\bibinfo {year} {2017})}\BibitemShut {NoStop}%
\bibitem [{\citenamefont {Cheng}\ and\ \citenamefont
  {Guo}(2017)}]{arXiv:1711.04408}%
  \BibitemOpen
  \bibfield  {author} {\bibinfo {author} {\bibfnamefont {J.~L.}\ \bibnamefont
  {Cheng}}\ and\ \bibinfo {author} {\bibfnamefont {C.}~\bibnamefont {Guo}},\
  }\href@noop {} {\bibfield  {journal} {\bibinfo  {journal} {arXiv:1711.04408}\
  } (\bibinfo {year} {2017})}\BibitemShut {NoStop}%
\bibitem [{\citenamefont {Tokman}\ \emph {et~al.}(2013)\citenamefont {Tokman},
  \citenamefont {Yao},\ and\ \citenamefont
  {Belyanin}}]{Phys.Rev.Lett._110_077404_2013_Tokman}%
  \BibitemOpen
  \bibfield  {author} {\bibinfo {author} {\bibfnamefont {M.}~\bibnamefont
  {Tokman}}, \bibinfo {author} {\bibfnamefont {X.}~\bibnamefont {Yao}}, \ and\
  \bibinfo {author} {\bibfnamefont {A.}~\bibnamefont {Belyanin}},\ }\href
  {\doibase 10.1103/PhysRevLett.110.077404} {\bibfield  {journal} {\bibinfo
  {journal} {Phys. Rev. Lett.}\ }\textbf {\bibinfo {volume} {110}},\ \bibinfo
  {pages} {077404} (\bibinfo {year} {2013})}\BibitemShut {NoStop}%
\bibitem [{\citenamefont {Shiri}\ and\ \citenamefont
  {Malakzadeh}(2017)}]{LaserPhys._27_16201_2017_Shiri}%
  \BibitemOpen
  \bibfield  {author} {\bibinfo {author} {\bibfnamefont {J.}~\bibnamefont
  {Shiri}}\ and\ \bibinfo {author} {\bibfnamefont {A.}~\bibnamefont
  {Malakzadeh}},\ }\href {http://stacks.iop.org/1555-6611/27/i=1/a=016201}
  {\bibfield  {journal} {\bibinfo  {journal} {Laser Phys.}\ }\textbf {\bibinfo
  {volume} {27}},\ \bibinfo {pages} {016201} (\bibinfo {year}
  {2017})}\BibitemShut {NoStop}%
\bibitem [{\citenamefont {Yang}\ \emph {et~al.}(2017)\citenamefont {Yang},
  \citenamefont {Chen}, \citenamefont {Xie}, \citenamefont {Liu},\ and\
  \citenamefont {Liu}}]{Sci.Rep._7_2513_2017_Yang}%
  \BibitemOpen
  \bibfield  {author} {\bibinfo {author} {\bibfnamefont {W.-X.}\ \bibnamefont
  {Yang}}, \bibinfo {author} {\bibfnamefont {A.-X.}\ \bibnamefont {Chen}},
  \bibinfo {author} {\bibfnamefont {X.-T.}\ \bibnamefont {Xie}}, \bibinfo
  {author} {\bibfnamefont {S.}~\bibnamefont {Liu}}, \ and\ \bibinfo {author}
  {\bibfnamefont {S.}~\bibnamefont {Liu}},\ }\href
  {https://doi.org/10.1038/s41598-017-02740-x} {\bibfield  {journal} {\bibinfo
  {journal} {Sci. Rep.}\ }\textbf {\bibinfo {volume} {7}},\ \bibinfo {pages}
  {2513} (\bibinfo {year} {2017})}\BibitemShut {NoStop}%
\bibitem [{\citenamefont
  {Solookinejad}(2016)}]{Phys.B_497_67_2016_Solookinejad}%
  \BibitemOpen
  \bibfield  {author} {\bibinfo {author} {\bibfnamefont {G.}~\bibnamefont
  {Solookinejad}},\ }\href {\doibase
  https://doi.org/10.1016/j.physb.2016.06.014} {\bibfield  {journal} {\bibinfo
  {journal} {Phys. B}\ }\textbf {\bibinfo {volume} {497}},\ \bibinfo {pages}
  {67 } (\bibinfo {year} {2016})}\BibitemShut {NoStop}%
\bibitem [{\citenamefont {Hamedi}\ and\ \citenamefont
  {Asadpour}(2015)}]{J.Appl.Phys._117_183101_2015_Hamedi}%
  \BibitemOpen
  \bibfield  {author} {\bibinfo {author} {\bibfnamefont {H.~R.}\ \bibnamefont
  {Hamedi}}\ and\ \bibinfo {author} {\bibfnamefont {S.~H.}\ \bibnamefont
  {Asadpour}},\ }\href {\doibase 10.1063/1.4919893} {\bibfield  {journal}
  {\bibinfo  {journal} {J. Appl. Phys.}\ }\textbf {\bibinfo {volume} {117}},\
  \bibinfo {pages} {183101} (\bibinfo {year} {2015})}\BibitemShut {NoStop}%
\bibitem [{\citenamefont {Shirley}(1965)}]{Phys.Rev._138_B979_1965_Shirley}%
  \BibitemOpen
  \bibfield  {author} {\bibinfo {author} {\bibfnamefont {J.~H.}\ \bibnamefont
  {Shirley}},\ }\href {\doibase 10.1103/PhysRev.138.B979} {\bibfield  {journal}
  {\bibinfo  {journal} {Phys. Rev.}\ }\textbf {\bibinfo {volume} {138}},\
  \bibinfo {pages} {B979} (\bibinfo {year} {1965})}\BibitemShut {NoStop}%
\bibitem [{\citenamefont {L\'opez-Rodr\'{\i}guez}\ and\ \citenamefont
  {Naumis}(2008)}]{Phys.Rev.B_78_201406_2008_Lopez-Rodriguez}%
  \BibitemOpen
  \bibfield  {author} {\bibinfo {author} {\bibfnamefont {F.~J.}\ \bibnamefont
  {L\'opez-Rodr\'{\i}guez}}\ and\ \bibinfo {author} {\bibfnamefont {G.~G.}\
  \bibnamefont {Naumis}},\ }\href {\doibase 10.1103/PhysRevB.78.201406}
  {\bibfield  {journal} {\bibinfo  {journal} {Phys. Rev. B}\ }\textbf {\bibinfo
  {volume} {78}},\ \bibinfo {pages} {201406} (\bibinfo {year} {2008})}
  \bibinfo {note} {; Phys. Rev. B {\bf 79}, 049901 (2009).}\BibitemShut {Stop}%
\bibitem [{\citenamefont {Calvo}\ \emph {et~al.}(2011)\citenamefont {Calvo},
  \citenamefont {Pastawski}, \citenamefont {Roche},\ and\ \citenamefont
  {Torres}}]{Appl.Phys.Lett._98_232103_2011_Calvo}%
  \BibitemOpen
  \bibfield  {author} {\bibinfo {author} {\bibfnamefont {H.~L.}\ \bibnamefont
  {Calvo}}, \bibinfo {author} {\bibfnamefont {H.~M.}\ \bibnamefont
  {Pastawski}}, \bibinfo {author} {\bibfnamefont {S.}~\bibnamefont {Roche}}, \
  and\ \bibinfo {author} {\bibfnamefont {L.~E. F.~F.}\ \bibnamefont {Torres}},\
  }\href {\doibase 10.1063/1.3597412} {\bibfield  {journal} {\bibinfo
  {journal} {Appl. Phys. Lett.}\ }\textbf {\bibinfo {volume} {98}},\ \bibinfo
  {pages} {232103} (\bibinfo {year} {2011})}\BibitemShut {NoStop}%
\bibitem [{\citenamefont {Sentef}\ \emph {et~al.}(2015)\citenamefont {Sentef},
  \citenamefont {Claassen}, \citenamefont {Kemper}, \citenamefont {Moritz},
  \citenamefont {Oka}, \citenamefont {Freericks},\ and\ \citenamefont
  {Devereaux}}]{Nat.Commun._6_7047_2015_Sentef}%
  \BibitemOpen
  \bibfield  {author} {\bibinfo {author} {\bibfnamefont {M.}~\bibnamefont
  {Sentef}}, \bibinfo {author} {\bibfnamefont {M.}~\bibnamefont {Claassen}},
  \bibinfo {author} {\bibfnamefont {A.}~\bibnamefont {Kemper}}, \bibinfo
  {author} {\bibfnamefont {B.}~\bibnamefont {Moritz}}, \bibinfo {author}
  {\bibfnamefont {T.}~\bibnamefont {Oka}}, \bibinfo {author} {\bibfnamefont
  {J.}~\bibnamefont {Freericks}}, \ and\ \bibinfo {author} {\bibfnamefont
  {T.}~\bibnamefont {Devereaux}},\ }\href
  {http://dx.doi.org/10.1038/ncomms8047} {\bibfield  {journal} {\bibinfo
  {journal} {Nat. Commun.}\ }\textbf {\bibinfo {volume} {6}},\ \bibinfo {pages}
  {7047} (\bibinfo {year} {2015})}\BibitemShut {NoStop}%
\bibitem [{\citenamefont {Cayssol}\ \emph {et~al.}(2013)\citenamefont
  {Cayssol}, \citenamefont {Dóra}, \citenamefont {Simon},\ and\ \citenamefont
  {Moessner}}]{PhysStatusSolidiRRL_7_101_2013_Cayssol}%
  \BibitemOpen
  \bibfield  {author} {\bibinfo {author} {\bibfnamefont {J.}~\bibnamefont
  {Cayssol}}, \bibinfo {author} {\bibfnamefont {B.}~\bibnamefont {Dóra}},
  \bibinfo {author} {\bibfnamefont {F.}~\bibnamefont {Simon}}, \ and\ \bibinfo
  {author} {\bibfnamefont {R.}~\bibnamefont {Moessner}},\ }\href {\doibase
  10.1002/pssr.201206451} {\bibfield  {journal} {\bibinfo  {journal} {Phys
  Status Solidi - Rapid research letters}\ }\textbf {\bibinfo {volume} {7}},\
  \bibinfo {pages} {101} (\bibinfo {year} {2013})}\BibitemShut {NoStop}%
\bibitem [{\citenamefont {Xu}\ \emph {et~al.}(2010)\citenamefont {Xu},
  \citenamefont {Sultan}, \citenamefont {Zhang},\ and\ \citenamefont
  {Cao}}]{Appl.Phys.Lett._97_11907_2010_Xu}%
  \BibitemOpen
  \bibfield  {author} {\bibinfo {author} {\bibfnamefont {X.~G.}\ \bibnamefont
  {Xu}}, \bibinfo {author} {\bibfnamefont {S.}~\bibnamefont {Sultan}}, \bibinfo
  {author} {\bibfnamefont {C.}~\bibnamefont {Zhang}}, \ and\ \bibinfo {author}
  {\bibfnamefont {J.~C.}\ \bibnamefont {Cao}},\ }\href {\doibase
  10.1063/1.3462972} {\bibfield  {journal} {\bibinfo  {journal} {Appl. Phys.
  Lett.}\ }\textbf {\bibinfo {volume} {97}},\ \bibinfo {pages} {011907}
  (\bibinfo {year} {2010})}\BibitemShut {NoStop}%
\bibitem [{\citenamefont {Zhou}\ and\ \citenamefont
  {Wu}(2011)}]{Phys.Rev.B_83_245436_2011_Zhou}%
  \BibitemOpen
  \bibfield  {author} {\bibinfo {author} {\bibfnamefont {Y.}~\bibnamefont
  {Zhou}}\ and\ \bibinfo {author} {\bibfnamefont {M.~W.}\ \bibnamefont {Wu}},\
  }\href {\doibase 10.1103/PhysRevB.83.245436} {\bibfield  {journal} {\bibinfo
  {journal} {Phys. Rev. B}\ }\textbf {\bibinfo {volume} {83}},\ \bibinfo
  {pages} {245436} (\bibinfo {year} {2011})}\BibitemShut {NoStop}%
\bibitem [{\citenamefont {Foa~Torres}\ \emph {et~al.}(2014)\citenamefont
  {Foa~Torres}, \citenamefont {Perez-Piskunow}, \citenamefont {Balseiro},\ and\
  \citenamefont {Usaj}}]{Phys.Rev.Lett._113_266801_2014_FoaTorres}%
  \BibitemOpen
  \bibfield  {author} {\bibinfo {author} {\bibfnamefont {L.~E.~F.}\
  \bibnamefont {Foa~Torres}}, \bibinfo {author} {\bibfnamefont {P.~M.}\
  \bibnamefont {Perez-Piskunow}}, \bibinfo {author} {\bibfnamefont {C.~A.}\
  \bibnamefont {Balseiro}}, \ and\ \bibinfo {author} {\bibfnamefont
  {G.}~\bibnamefont {Usaj}},\ }\href {\doibase 10.1103/PhysRevLett.113.266801}
  {\bibfield  {journal} {\bibinfo  {journal} {Phys. Rev. Lett.}\ }\textbf
  {\bibinfo {volume} {113}},\ \bibinfo {pages} {266801} (\bibinfo {year}
  {2014})}\BibitemShut {NoStop}%
\bibitem [{\citenamefont {Kundu}\ \emph {et~al.}(2014)\citenamefont {Kundu},
  \citenamefont {Fertig},\ and\ \citenamefont
  {Seradjeh}}]{Phys.Rev.Lett._113_236803_2014_Kundu}%
  \BibitemOpen
  \bibfield  {author} {\bibinfo {author} {\bibfnamefont {A.}~\bibnamefont
  {Kundu}}, \bibinfo {author} {\bibfnamefont {H.~A.}\ \bibnamefont {Fertig}}, \
  and\ \bibinfo {author} {\bibfnamefont {B.}~\bibnamefont {Seradjeh}},\ }\href
  {\doibase 10.1103/PhysRevLett.113.236803} {\bibfield  {journal} {\bibinfo
  {journal} {Phys. Rev. Lett.}\ }\textbf {\bibinfo {volume} {113}},\ \bibinfo
  {pages} {236803} (\bibinfo {year} {2014})}\BibitemShut {NoStop}%
\bibitem [{\citenamefont {Jauho}\ and\ \citenamefont
  {Johnsen}(1996)}]{Phys.Rev.Lett._76_4576_1996_Jauho}%
  \BibitemOpen
  \bibfield  {author} {\bibinfo {author} {\bibfnamefont {A.~P.}\ \bibnamefont
  {Jauho}}\ and\ \bibinfo {author} {\bibfnamefont {K.}~\bibnamefont
  {Johnsen}},\ }\href {\doibase 10.1103/PhysRevLett.76.4576} {\bibfield
  {journal} {\bibinfo  {journal} {Phys. Rev. Lett.}\ }\textbf {\bibinfo
  {volume} {76}},\ \bibinfo {pages} {4576} (\bibinfo {year}
  {1996})}\BibitemShut {NoStop}%
\bibitem [{\citenamefont {Kibis}\ \emph {et~al.}(2016)\citenamefont {Kibis},
  \citenamefont {Morina}, \citenamefont {Dini},\ and\ \citenamefont
  {Shelykh}}]{Phys.Rev.B_93_115420_2016_Kibis}%
  \BibitemOpen
  \bibfield  {author} {\bibinfo {author} {\bibfnamefont {O.~V.}\ \bibnamefont
  {Kibis}}, \bibinfo {author} {\bibfnamefont {S.}~\bibnamefont {Morina}},
  \bibinfo {author} {\bibfnamefont {K.}~\bibnamefont {Dini}}, \ and\ \bibinfo
  {author} {\bibfnamefont {I.~A.}\ \bibnamefont {Shelykh}},\ }\href {\doibase
  10.1103/PhysRevB.93.115420} {\bibfield  {journal} {\bibinfo  {journal} {Phys.
  Rev. B}\ }\textbf {\bibinfo {volume} {93}},\ \bibinfo {pages} {115420}
  (\bibinfo {year} {2016})}\BibitemShut {NoStop}%
\bibitem [{\citenamefont {Torres}\ and\ \citenamefont
  {Kunold}(2005)}]{Phys.Rev.B_71_115313_2005_Torres}%
  \BibitemOpen
  \bibfield  {author} {\bibinfo {author} {\bibfnamefont {M.}~\bibnamefont
  {Torres}}\ and\ \bibinfo {author} {\bibfnamefont {A.}~\bibnamefont
  {Kunold}},\ }\href {\doibase 10.1103/PhysRevB.71.115313} {\bibfield
  {journal} {\bibinfo  {journal} {Phys. Rev. B}\ }\textbf {\bibinfo {volume}
  {71}},\ \bibinfo {pages} {115313} (\bibinfo {year} {2005})}\BibitemShut
  {NoStop}%
\bibitem [{Note1()}]{Note1}%
  \BibitemOpen
  \bibinfo {note} {Equation~(\ref {eq:quasi-eigenf}) gives $\DOTSB \sum@
  \slimits@ _{n_1n_2}eE^{(n_1);d}_{\protect \text {drv}}\protect \mathaccentV
  {hat}05E\xi ^du_\alpha ^{(n_2)}[u_\alpha ^{(n_1+n_2)}]^\protect \dag =\DOTSB
  \sum@ \slimits@ _l(l\hbar \Omega +\epsilon _\alpha -\protect \mathaccentV
  {hat}05EH_0)u_\alpha ^{(l)}[u_{\alpha }^{(l)}]^\protect \dag $. By
  substituting this expression into the numerator of Eq.~(\ref {eq:gl}), it can
  be simplified to $[\protect \mathaccentV {hat}05EH_0,\protect \mathaccentV
  {hat}05E\rho _0]$, which is zero.}\BibitemShut {Stop}%
\bibitem [{\citenamefont {Jiang}\ and\ \citenamefont
  {Wu}(2007)}]{Phys.Rev.B_75_035307_2007_Jiang}%
  \BibitemOpen
  \bibfield  {author} {\bibinfo {author} {\bibfnamefont {J.~H.}\ \bibnamefont
  {Jiang}}\ and\ \bibinfo {author} {\bibfnamefont {M.~W.}\ \bibnamefont {Wu}},\
  }\href {\doibase 10.1103/PhysRevB.75.035307} {\bibfield  {journal} {\bibinfo
  {journal} {Phys. Rev. B}\ }\textbf {\bibinfo {volume} {75}},\ \bibinfo
  {pages} {035307} (\bibinfo {year} {2007})}\BibitemShut {NoStop}%
\end{thebibliography}

\end{document}